\documentstyle{article}
\title{A nonlocal classical perspective on \\
         quantum electrodynamics}
\author{P. W. Morgan\\
        {\small\sl \vbox{\vskip 3pt
                         \hbox{30, Shelley Road,} 
                         \hbox{Oxford, OX4 3EB,} 
                         \hbox{England.}}}\\
        {\small\sl \hbox{peter.morgan@philosophy.oxford.ac.uk}}}
\date{September 2nd, 2001}
\def\Xabstract{The ideas behind the nonlocal classical statistical field
theory model for the quantized Klein-Gordon field introduced in Morgan(2001,
quant-ph/0106141) are extended to accommodate quantum electrodynamics. The
anticommutation rules for the quantized Dirac spinor field are given a
classical interpretation as a relativistically covariant modification of
the minimal coupling interaction between the classical electromagnetic field
and a classical Dirac spinor field.}

\begin{document}
\hoffset=-0.130\hsize
\hsize = 1.2\hsize

\def\Planck{{\vrule width 0.4em height 0.53em depth -0.5em\kern -0.45em h}}
\def\Half{{\scriptstyle {\scriptstyle 1 \over \scriptstyle 2}}}
\def\kT{{{\rm\sl k}T}}
\def\deltaKK{{2\pi\delta(k_\mu k^\mu\!-\!m^2)}}

\def\citeMorgan{{Morgan(2001)}}
\def\citeIZS221S331{{Itzykson and Zuber(1980, \S 2-2-1, \S 3-3-1)}}
\def\citeIZtable61{{Itzykson and Zuber(1980, table 6-1, p275)}}
\def\citeWeinbergP236{{Weinberg(1995, pp236-238)}}
\def\citeSWss{{Streater and Wightman(1964, Theorem 4-10)}}

\maketitle

\hsize=0.9\hsize
\hangindent=0.111\hsize\hangafter=-20
\Xabstract

\vskip 30pt
\hsize=1.1111\hsize
\baselineskip 16.00pt

In \citeMorgan, I constructed a relativistically nonlocal classical
statistical field theory hidden variable model for the quantized
Klein-Gordon field. Here I construct a relativistically nonlocal classical
statistical field theory hidden variable model for quantum electrodynamics
on similar lines. Such a model should not be taken to be how the world really
is, particularly because there is certainly no immediate extension of this
model to accommodate the whole of the standard model of particle physics, but
visualizability of a model is a pragmatic advantage, at least to me.

The approach of \citeMorgan\ does not yield a classically acceptable
statistical field theory model for interaction-free fermion fields,
specifically because of the anticommutation relations. We will therefore
take boson fields to be primary in pursuing a classical nonlocal model for
quantum electrodynamics, because we can construct a nonlocal classical
statistical field theory model for the interaction-free quantized
electromagnetic field by the same methods as were introduced in \citeMorgan.
We will find, at least for the purposes of perturbation theory, that by a
suitable adjustment elsewhere in the theory we can take Dirac spinor fields
effectively to be boson fields.

To construct a classical electrodynamics, we construct a nonlocal classical
statistical field theory model for the interaction-free electromagnetic
field (with a 2-point correlation function that is the same as the particle
propagator of the quantized electromagnetic field in quantum
electrodynamics), then we introduce a classical Dirac spinor field with
a 2-point correlation function that is the same as the particle propagator
of the quantized Dirac spinor field in quantum electrodynamics (which we can
do, but there will only be very distant relationships between probability
measures over classical field values and probability measures over
observables of the quantum field). We also introduce the same minimal
coupling interaction of the Dirac spinor field with the electromagnetic
field as occurs in quantum electrodynamics. In such a classical theory, we
can in principle eliminate the classical Dirac spinor field to give a system
of nonlinear, higher order equations for the electromagnetic field alone.

The Feynman rules for perturbation expansions in this classical statistical
field theory model are then identical in our classical electrodynamics to
the Feynman rules for perturbation expansions in quantum electrodynamics,
except for the sign switching rules (see, for example, rule 4 in
\citeIZtable61): a minus sign must be introduced for every closed fermion
loop, and a minus sign must be introduced for odd permutations of external
fermion lines. These sign switching rules are the sole consequence in the
Feynman rules of the anticommutation properties of quantized Dirac spinor
fields. Although the anticommutation is essential in quantum field theory
to ensure relativistic signal locality in quantum field theory (see, for
example, \citeWeinbergP236, or \citeSWss), from a classical point of view
it is far more important that these sign switching rules are required for
empirical accuracy. It is reasonable to introduce such sign switching rules
just to achieve empirical accuracy.
 
If we consider ourselves to be constructing a nonlinear field theory for
the electromagnetic field, then the only classical Feynman diagrams we are
really interested in have external lines only for the electromagnetic field,
so if we can justify the first rule, then the second rule can be taken to be
necessary for consistency. In the context of quantum field theory, it has 
been taken that the sign switching of the interaction terms is a property
of the quantized Dirac spinor field, because anticommutation is required for
relativistic signal locality even for the interaction-free quantized Dirac
spinor field. In a classical context, however, the sign switching of the
interaction terms can be taken to be a property of the interaction terms,
because there is no equivalent formal requirement on interaction-free
classical Dirac spinor fields (indeed, it is quite common to refer to
``exchange forces'' in particle physics, which implicitly takes the sign
switches to be a property of the interaction rather than of the fields).

On this classical view, the interaction between the interaction-free
classical fields is not just the relativistically covariant minimal coupling,
but is also the relativistically covariant specification of changes of sign
in the perturbation expansion. Note that for this modified minimal coupling
we can still in principle eliminate the classical Dirac spinor field to give
a (different) system of nonlinear, higher order equations for the
electromagnetic field.

For empiricists and post-empiricists this description of how to get the 
perturbation expansion empirically correct should be quite adequate,
since it is entirely through perturbation expansions that we usually take
experimental results to be adequately described by quantum electrodynamics.
For others, the beautiful mathematical structure of quantum field theory in
terms of the Wightman axioms, say, with its necessity of anticommutation
rules for quantized Dirac spinor fields, is unlikely to be supplanted by the
introduction in the classical context of the sign switching rules for the
interaction just to ensure empirical adequacy. There remains the possibility,
however, that the sign switching rules may be natural in a modified or
different classical formalism. With the availability of this type of
classical model for quantum electrodynamics, it in any case seems reasonable
enough to incorporate quantum field theory, and the naturalness of the sign
switching rules in quantum field theory, into classical physics. A
straightforward and conceptually conservative interpretation of quantum
field theory is possible if we think of it as no more than a particularly
effective calculational strategy for classical statistical field theory.

\vskip 50pt
\noindent{\Large\bf Bibliography}
\def\RefStart{\vskip 12.5pt\noindent\hangindent 30pt}

\baselineskip 13.5pt
\RefStart
   {Itzykson, C., and Zuber, J.-B. (1980),
       {\sl Quantum Field Theory}
       (New York: McGraw-Hill).}

\RefStart
   {Morgan, P. (2001),
       `A classical perspective on nonlocality in quantum field theory';
       quant-ph/0106141.} 

\RefStart
   {Streater, R. F., and Wightman, A. S. (1964),
       {\sl PCT, Spin and Statistics, and all that},
       (New York: W. A. Benjamin).}

\RefStart
   {Weinberg, S. (1995), {\sl The Quantum Thoery of Fields}, Volume I
       (Cambridge: Cambridge University Press).}

\end{document}